# Solving the disordered structure of β-$Cu_{2-x}Se$ using the three-dimensional difference pair distribution function

Authors

**Nikolaj Roth[a] and Bo B. Iversen[a]***

[a]Center for Materials Crystallography, Department of Chemistry, Aarhus University, Aarhus, 8000, Denmark

Correspondence email: bo@chem.au.dk

**Synopsis** Using three-dimensional difference pair distribution function analysis of single crystal diffuse X-ray scattering, the disordered structure copper selenide (β-$Cu_{2-x}Se$) at room temperature is solved. The structure is ordered in two dimensions but disordered in the third.

**Abstract** High-performing thermoelectric materials such as $Zn_4Sb_3$ and clathrates have atomic disorder as the root to their favourable properties. This makes it extremely difficult to understand and model their properties at a quantitative level, and thus effective structure-property relations are challenging to obtain. $Cu_{2-x}Se$ is an intensely studied, cheap and non-toxic high performance thermoelectric, which exhibits highly peculiar transport properties, especially near the β to α phase transition around 400 K, which must be related to the detailed nature of the crystal structure. Attempts to solve the crystal structure of the low temperature phase, β-$Cu_{2-x}Se$, have been unsuccessful since 1936. So far, all studies have assumed that β-$Cu_{2-x}Se$ has a three-dimensional periodic structure, but here we show that the structure is ordered only in two dimensions while it is disordered in the third dimension with a near random stacking sequence. Using the three-dimensional difference pair distribution function (3D-ΔPDF) analysis method for diffuse single crystal X-ray scattering, we solve the structure of the ordered layer and show that there are two modes of stacking disorder present, which give rise to an average structure with higher symmetry. The present approach allows for a direct solution of structures with disorder in some dimensions and order in others, and can be though of as a generalization of the crystallographic Patterson method. The local and extended structure of a solid determines its properties and $Cu_{2-x}Se$ represents an example of a high-performing thermoelectric material where the local atomic structure differs significantly from the average periodic structure observed from Bragg crystallography.

## 1. Introduction

The discovery of disorder effects in crystal structures, which were previously believed to be ordered, has made it possible to explain properties of solids, which could not previously be understood. This has been especially important to the field of thermoelectrics, where disorder effects have clarified the fundamental origin of the low thermal conductivity in e.g. clathrates and $Zn_4Sb_3$ (2018; Christensen *et al.*, 2010; Snyder *et al.*, 2004). Ordered crystal structures are solved through analysis of either X-ray or neutron diffraction data, where the periodicity of the crystal leads to sharp and intense Bragg peaks in the diffraction pattern. Disorder on the other hand gives rise to weak diffuse scattering, which is broad in reciprocal space. In crystals with an average order and local disorder, the scattering will contain both sharp Bragg peaks as well as weak diffuse scattering, making it difficult to measure the diffuse scattering accurately, as it is often more than three orders of magnitude weaker than the Bragg diffraction. In recent years, the measurement of diffuse scattering from single crystals has been made easier due to improvements in detector technology as well as improved X-ray and neutron sources.

$Cu_{2-x}Se$ is a promising thermoelectric material as it is cheap, nontoxic and has a high thermoelectric figure of merit (zT) (Eikeland *et al.*, 2017; Liu, Yuan, *et al.*, 2013; Lu *et al.*, 2015; Qiu, 2016; Nguyen *et al.*, 2013; Brown *et al.*, 2016; Brown, Day, Caillat, *et al.*, 2013; Brown, Day, Borup, *et al.*, 2013; Dalgaard *et al.*, 2018; Liu, Shi, *et al.*, 2013; Liu *et al.*, 2012; Miyatani, 1973; Vučić *et al.*, 1984; Vucic *et al.*, 1982; Vučić *et al.*, 1981; Yu *et al.*, 2012; Chi *et al.*, 2014; Mahan, 2015; Sirusi *et al.*, 2015; Sun *et al.*, 2015; Stephen Dongmin *et al.*, 2016). The compound has a phase transition at 350-410 K depending on the degree of Cu deficiency (Brown, Day, Caillat, *et al.*, 2013; Chi *et al.*, 2014; Eikeland *et al.*, 2017; Liu, Shi, *et al.*, 2013; Liu, Yuan, *et al.*, 2013; Vučić *et al.*, 1984). Both low- and high-temperature phases have been studied for their thermoelectric properties, while the high-temperature phase is also a superionic conductor (Miyatani, 1973; Vucic *et al.*, 1982; Mahan, 2015; Sirusi *et al.*, 2015; Stephen Dongmin *et al.*, 2016). The compound has been described as a phonon-liquid electron-crystal material due to its good electrical conductivity and low thermal conductivity (Liu *et al.*, 2012). Much interest has concerned the highly peculiar behaviour of the transport properties around the phase transition, where a large enhancement in zT is observed (Brown, Day, Borup, *et al.*, 2013; Brown *et al.*, 2016; Liu, Yuan, *et al.*, 2013).

There is some confusion in the literature about the nomenclature for the low- and high-temperature phase of $Cu_{2-x}Se$. Here we use the original labelling by Rahlfs (Rahlfs, 1936), which names the low-temperature phase as β-$Cu_{2-x}Se$ and the high temperature phase as α-$Cu_{2-x}Se$. The high temperature α-$Cu_{2-x}Se$ has the cubic anti-fluorite structure with the Cu atoms distributed in a disordered way on the corners and center of tetrahedra between the layers of the fcc Se-lattice (Dalgaard *et al.*, 2018; Eikeland *et al.*, 2017; Danilkin, 2011, 2012).

The structure of β-$Cu_{2-x}Se$ has been discussed since 1936 (Rahlfs, 1936), and more than 10 different unit cells for the structure have been proposed, some of which were collected in the 1987 paper by Milat et al. (Milat *et* *al.*, 1987). The crystal systems of these proposed unit cells span triclinic, monoclinic, orthorhombic, tetragonal, trigonal and cubic. In 2011 Gulay *et al.* reported to have solved the structure of β-$Cu_{2-x}Se$ from X-ray diffraction data (Gulay *et al.*, 2011), but their model had a very low agreement with the data and was later shown to be incorrect (Eikeland *et al.*, 2017). Later studies by Brown *et al.* attempted to find the structure using pair distribution function and density functional theory analysis (Brown, Day, Borup, *et al.*, 2013; Brown, Day, Caillat, *et al.*, 2013; Brown *et al.*, 2016). Even with all these large efforts in understanding the structure of β-$Cu_{2-x}Se$, there is still a wide acknowledgement that a sufficient structural model has not yet been found (Brown *et al.*, 2016; Dalgaard *et al.*, 2018; Eikeland *et al.*; 2017, Frangis *et al.*, 1991, Gulay *et al.*, 2011, Liu, Yuan, *et al.*, 2013; Lu *et al.*, 2015; Milat *et al.*, 1987; Nguyen *et al.*, 2013; Yamamoto & Kashida, 1991; Brown, Day, Caillat, *et al.*, 2013; Kashida & Akai, 1988; Qiu, 2016; Rahlfs, 1936). One reason for this lack of structural information is the difficulty in synthesizing high quality single crystals of the β phase. As the high temperature α phase is cubic, the low temperature phase easily becomes twinned. However, this difficulty is not enough to explain the problem of solving the structure. In 2017 Eikeland *et al.* showed that the diffraction pattern of a very high quality single crystal of β-$Cu_{2-x}Se$ can be divided into strong main reflections and weak superstructure-reflections and was able to solve the average structure from the main peak intensities only (Eikeland *et al.*, 2017). This average structure contains ordered Se sites and Cu on disordered sites. The information about the Cu ordering is then contained in the weak reflections.

Figure 1a shows the HK0 plane of the scattered X-ray intensity from β-$Cu_{2-x}Se$ at 300K, where the strong main reflections and weak superstructure reflections can be seen. Several superstructures for the Cu ordering have been proposed, of which most have very small differences in energy (Eikeland *et al.*, 2017; Gulay *et al.*, 2011, Liu, Yuan, *et al.*, 2013; Lu *et al.*, 2015; Nguyen *et al.*, 2013; Qiu, 2016). However, none of these have so far been in good agreement with experimental X-ray scattering data. We show here that the previously reported superstructure reflections are not Bragg reflections, but instead diffuse scattering rods, showing the structure to have a two-dimensional ordered superstructure and one-dimensional disorder. These diffuse rods along the l-direction of reciprocal space can be seen in Figure 1b-c. We analyse this diffuse scattering data using the three-dimensional difference pair distribution function (3D-ΔPDF) method (Canut-Amorös, 1967; Schaub *et al.*, 2007; Weber *et al.*, 2007; Weber & Simonov, 2012), which makes it possible to directly construct the two-dimensional superstructure of the layers as well as identify the one-dimensional disorder of the layer stacking. The 3D-ΔPDF is defined as the inverse Fourier transform of the scattered diffuse intensity, which is equal to the autocorrelation of the difference between the total electron density and the average periodic electron density (Weber & Simonov, 2012):

$$\text{3D-}\Delta\text{PDF} = \mathcal{F}^{-1}\left[I_{diffuse}\right] = \langle \delta\rho \otimes \delta\rho \rangle$$

Where, $\delta\rho(\boldsymbol{r}) = \rho_{total}(\boldsymbol{r}) - \rho_{periodic}(\boldsymbol{r})$ is the difference electron density, $\langle \dots \rangle$ is the experiment time-average and $\otimes$ the cross-correlation operator. The autocorrelation of the difference density will have positive peaks for vectors separating more electron density than in the average periodic structure, and negative peaks for vectors separating less electron density than the average periodic structure. As the 3D-ΔPDF gives a direct-space three-dimensional view of the correlations in the system, it provides an intuitive way to understand the information contained in the scattering data. It can be thought of as a generalized Patterson function for diffuse scattering. Unlike the frequently used one-dimensional PDF technique (Billinge & Egami, 2003), the 3D-ΔPDF method separates interactions at equal distances but different spatial directions, and it also makes observation of weak disorder possible in systems with a superimposed average order.

From the 3D-ΔPDF analysis, we find that the structure consists of ordered layers, which are stacked in a disordered way. There is only one type of layer, which can be described using layer group symmetry, whereby only coordinates for five atoms are needed. We then find that there is a disordered stacking sequence of this layer and its mirror image as well as a disorder in three possible inter-layer vectors. We then compare the found structure to suggested structures in the literature and show that many of the previously proposed structures are periodic repetitions of units, which will be found in the real disordered structure.

## 2. Methods

Single crystals of β-$Cu_{2-x}Se$ were grown using chemical vapour transport (CVT) with iodine as a transport agent as previously reported in (Eikeland *et al.*, 2017). The average structure was studied at 300 K on a SuperNova diffractometer from Agilent Technologies, using Mo Kα radiation (λ= 0.71073 Å). Diffracted intensities were collected on a CCD detector and the data integrated and corrected for absorption using CrysAlisPro (Xcalibur, 2010). The structure solution and refinement were carried out with SHELXS and SHELXL, respectively, using the Olex2 GUI (Dolomanov *et al.*, 2009; Sheldrick, 2015; Sheldrick, 2008). The occupancies of all Cu sites were allowed to refine freely, while the occupancy of Se was fixed to 1. The diffuse scattering experiment was carried out at the 15-ID-D beamline at the Advanced Photon Source. A single crystal of $Cu_{2-x}Se$ (90 μm octahedron) was glued to the end of a thin glass pin using epoxy and mounted on a Huber kappa-geometry goniometer. Data was measured using 40 keV X-rays with a sample to detector distance of 120 mm on a Dectris Pilatus 1M CdTe detector. As the detector has gaps between the detector modules, 3 runs were measured with different detector positions to avoid any missing data. Each run was a 360 degree omega rotation with phi and kappa at 0. Each frame was measured every 0.1 degrees during a continuous rotation. After the experiment the crystal was removed and the background air scattering was measured separately for later subtraction. The data was converted to reciprocal space using a Matlab script written by Martin von Zimmerman (Zimmermann, 2019). During this process the data was corrected for Lorentz and polarization factors, the background scattering from air was subtracted, and a solid angle correction was applied as the detector is flat. The resulting data was symmetrized using the $\bar{3}m$ point symmetry of the Laue group. The resulting scattering data was reconstructed on a 901 x 901 x 901 point grid with each axis spanning ±19.5 $Å^{-1}$.

In order to obtain the 3D-ΔPDF, the Bragg peaks were removed and replaced by interpolated values using a punch and fill method, as previously described in the supporting information of Roth *et al.* (2018). The resulting data containing only the diffuse scattering was then Fourier transformed to give the 3D-ΔPDF.

The theoretical scattering pattern based on the found model for the structure was calculated using a python script. First the atomic scattering factors for Cu and Se are calculated on a 501 x 501 x 501 point grid for the same range as the experimental data using the parametrized scattering factors by Doyle and Turner (Doyle & Turner, 1968). The scattering for one unit cell of the ordered layer structure is then calculated using the structure factor equation $F_{cell}(\boldsymbol{q}) = \sum_j f_j(\boldsymbol{q}) \exp(i\boldsymbol{r}_j\boldsymbol{q})$ where $\boldsymbol{q}$ is the scattering vector and $\boldsymbol{r}_j$ the coordinate of the j'th atom in the layer unit cell based on the model in Table 2. The scattering of one ordered layer is calculated by $F_{layer}(\boldsymbol{q}) = \sum_j F_{cell}(\boldsymbol{q}) \exp(i\boldsymbol{t}_j\boldsymbol{q})$ where $\boldsymbol{t}_j$ is the translational vector to the j'th unit cell in two dimensions. The scattering of the mirrored layers is given by the mirror of $F_{layer}$. The scattering for a sequence of layers is then calculated as $F_{model}(\boldsymbol{q}) = \sum_j F_{layer_j}(\boldsymbol{q}) \exp(i\boldsymbol{s}_j\boldsymbol{q})$ where $\boldsymbol{s}_j$ is the vector to the j'th layer and $F_{layer_j}$ is the scattering factor for the j'th layer. Finally the intensity is obtained through $I = |F_{model}|^2$. To calculate the 3D-ΔPDF based on the found model, the calculated intensity for the average structure is subtracted from the calculated intensity for the model and the Fourier transform is taken.

## 3. Average structure

The average structure which was solved and refined from the Bragg diffraction data is identical to the one previously reported (Eikeland *et al.*, 2017) and is shown in Figure 2. Structural data is given in Table 1. The structure is build up from hexagonally close packed Se layers stacked in a distorted face centered cubic (fcc) pattern along the *c*-axis in the hexagonal unit cell. The Cu atoms are disordered over three different sites. Two of these sites, marked Cu1a and Cu1b in Figure 2a are very close and have occupancies of 1/3 and 2/3 respectively, within two standard deviations. From chemical considerations, it is expected that when a Cu1a site is filled, the Cu1b right next to it will be empty and vice versa. The last Cu site, marked Cu2, forms small triangles with each corner filled 1/3 as seen in Figure 2b. Also for this site it is expected that only one Cu position in each of the triangles will be occupied in the real structure. The structure can be seen as layered, where each layer consist of two Se layers, two Cu1a layers, two Cu1b layers and two Cu2 layers. One such layer is marked with a curly bracket in Figure 2a, and Figure 2b shows one such layer from above. Between these layers are gaps of ~3Å, and the interactions between the layers is expected to be weak. From the structural refinement, the composition is found to be $Cu_{1.95(2)}Se$, in quantitative agreement with refinements on previous crystal from the same synthesis batch (Eikeland *et al.*, 2017), which suggests that there is a ~2.5 % deficiency of Cu. However, it should be mentioned that the present refinements, like the previous in the literature, used the independent atom model with neutral Cu and Se species. Since the structure probably contains species closer to $Cu^+$ and $Se^{2-}$, the real Cu content may in fact be closer to the stoichiometric $Cu_2Se$.

#### 4. Two-dimensional order - one-dimensional disorder

The diffuse scattering contains the information on the local deviations from the average structure. In Figure 1b and c strong diffuse features are seen as thin lines along the $l$-direction of reciprocal space. This one-dimensional diffuse scattering reveals the presence of two-dimensional ordering and one-dimensional disorder. From the average structure, it is expected that the one-dimensional disorder may arise from stacking faults of layers, which have a two-dimensional ordered superstructure. To confirm this and to find the exact ordering and disorder we use the 3D-ΔPDF, which is obtained as the Fourier transform of the diffuse scattering. The amplitude of a peak in the 3D-ΔPDF for a vector $\boldsymbol{r}$ is given by $\sum_{(i,j)|\boldsymbol{r}_{ij}=\boldsymbol{r}} \delta Z_i \, \delta Z_j$ where $\delta Z_i$ is the difference in number of electrons between the real and average structure for atom i (see supporting information for more details). The summation runs over all pairs of atoms (i, j) separated by the same vector $\mathbf{r}$.

To illustrate the method by which the 2D ordered structure can be solved, a layer of Cu2 sites is shown in Figure 3 together with a small part of the 3D-ΔPDF for the z = 0 plane. The Cu2 layers are perpendicular to the *c*-direction of the unit cell of the average structure and contain small triangles of Cu, with each corner of the triangles having 1/3 occupancy in the average structure.

We start by assuming that the site marked “O” is occupied by a Cu atom. The two immediate neighbors in the same triangle are separated from “O” by a distance of 0.88 Å and cannot be occupied at the same time as “O”. This is confirmed by the 3D-ΔPDF, where there is a positive peak at the origin surrounded immediately by a negative ring. Positive peaks in the 3D-ΔPDF show vectors for which the real structure has more electron density separated by that vector compared with the average structure. In the same way a negative peak occurs for vectors which separate less electron density compared with the average structure. As all atoms are always separated from themselves by the zero vector, which is different to the average structure, a positive peak will always be found at the zero vector in the 3D-ΔPDF. The average structure gives electron density separated by the vectors between the Cu sites within one triangle. As this does not occur in the real structure, these vectors shows negative regions in the 3D-ΔPDF. To find the ordering of Cu between the triangles we then look at longer vectors in the 3D-ΔPDF. Some of the sites in Figure 3 are marked by numbers. The vectors from “O” to the numbered sites are marked in the 3D-ΔPDF using the same numbers. It is found that for vectors from “O” to each corner of a triangle, only one will be in a positive region of the 3D-ΔPDF, and the remaining two in the negative regions. This shows that there is a unique ordering of the Cu2 layers. The red-tinted sites in Figure 3 show the occupied Cu sites assuming that “O” is occupied. Longer correlations can then be seen in Figure 4, where larger cuts through the 3D-ΔPDF are shown. The same method for finding the ordered structure of a layer with the same z-coordinate is then used for the other Cu layers in the structure. The correlations between layers are then found for the z ≠ 0 Å parts of the 3D-ΔPDF. Figure 4b shows one such layer with z = 0.97 Å. In this way, the ordered superstructure can be solved, as the possible sites and amounts of filling is known from the average structure. A detailed step-by-step guide through the structure solution is given in the supporting information.

The one-dimensional disorder can also be seen from the 3D-ΔPDF when looking at longer correlations in the z-direction. Figure 4c shows the y = 0 Å plane. The ordering of the layers is seen as strong features in the xy plane, while the stacking disorder is seen by the fast decay of features in the z direction. The longest distance in the z-direction with strong features is seen to be around 5Å. This is also the thickness of the layers in the structure and thus there is disordering in the stacking of layers. It is therefore meaningful to describe the structure by perfectly ordered two-dimensional layers with stacking disorder.

The found ordered layer superstructure is shown in Figure 5. The layer is three-dimensional with translational symmetry in two dimensions and point symmetry in the third. This type of symmetry is best described using the subperiodic layer groups (Kopsky & Litvin, 2002), and the layer has $p\bar{3}$ layer group symmetry. The new unit cell and symmetry elements of the layer are shown in Figure 5b. Using this description only one Se site and four Cu sites are needed. The unit cell dimensions, atomic coordinates, Wyckoff positions and their related sites in the average structure are given in Table 2. Note that the coordinates x and y are given as relative coordinates with respect to the unit cell axis, while the z coordinate is given in Å. In the supporting information we show that the 3D-ΔPDF requires two types of layers, which have the same structure but are related by the mirror operation (x,y,z)→(y,x,z) in the layer coordinate system. As the layer itself has an inversion center this mirror is the same as turning the layer upside-down by rotating the layer 180 degrees around $[1\bar{1}0]$.

The one-dimensional disorder along the *c*-direction can be explained through the symmetry relation between the average structure and the ordered layer structure. The unit cell of the ordered layer structure is a superstructure in the $ab$ plane with $\boldsymbol{a}_{layer} = \boldsymbol{a}_{avg} - \boldsymbol{b}_{avg}$ and $\boldsymbol{b}_{layer} = 2\boldsymbol{b}_{avg} + \boldsymbol{a}_{avg}$, which makes $|\boldsymbol{a}_{layer}| = \sqrt{3}|\boldsymbol{a}_{avg}|$, which we show in the supporting information. As the ordered layer symmetry is $p\bar{3}$ while the average structure has symmetry $R\bar{3}m$, the disorder in stacking gives the average structure its mirror plane and the R-centering as well as a smaller unit cell. The mirror plane gives the possibility of two types of layers in the structure, one which is the mirror image of the other, which was also identified from the 3D-ΔPDF (see the supporting information). In the average structure the relation from one layer to the next is given by the R-centering vector $(^2/_3, ^1/_3, ^1/_3)$ which by symmetry is equivalent to $(^{-1}/_3, ^1/_3, ^1/_3)$ and $(^{-1}/_3, ^{-2}/_3, ^1/_3)$. In the ordered layer structure, these are no longer equivalent because of the lower symmetry. In the coordinate system of the ordered layer these vectors are $(^{-1}/_3, 0, 6.81\text{Å})$ , $(0, ^{-1}/_3, 6.81\text{Å})$ and $(^1/_3, ^1/_3, 6.81\text{Å})$. This means that there are two modes of disorder in the *c*-direction of $Cu_{2-x}Se$. One is whether each layer is mirrored or not and the other which of the three inter-layer vectors relate one layer to the next. Both of these disorder modes are needed to construct the average structure. From the 3D-ΔPDF along the *z*-direction, as seen in Figure 4c, there are no significant features at lengths corresponding to inter-layer vectors, and this means that the stacking of layers is highly disordered and nearly random.

To check the validity of the proposed model, the scattering and resulting 3D-ΔPDF has been calculated for a model with a random stacking of layers, where both the sequence of mirrored and non-mirrored layers is random as well as a random order of the three inter-layer vectors connecting one layer to the next. The 3D-ΔPDF from this model is shown in Figure 4d, e and f, and a good agreement with the measured 3D-ΔPDF is found. The calculated scattering for three reciprocal space planes is shown together with the experimental data in Figure 1. Both the strong Bragg peaks and thin diffuse lines are recreated in the scattering pattern calculated from the model. Theoretical scattering patterns were also calculated from structural models where only one of the modes of disorder (mirror or inter-layer translation) were used, but these patterns are not in agreement with experimental data (shown in the supporting information). When comparing the experimental and calculated scattering, it is seen that although both the strong Bragg peaks and weak diffuse scattering rods are reproduced by the calculated model, there are a few very weak features in the experiment which are not reproduced by the model. In the experiment a few very weak peaks can be found halfway between some of the Bragg peaks in the $q_z$-direction. This suggest that there is a small modulation of the average structure, doubling the c-axis. However these peaks are around three orders of magnitude weaker than Bragg peaks, showing this modulation to be very small. Likewise the diffuse scattering rods in the experiment show a slight modulation of intensity along $q_z$ not seen in the calculated data, suggesting that there might be weak correlations in the highly disordered stacking order of layers. The experimental data also shows a very weak broad diffuse scattering term, which is not reproduced by the calculation. This diffuse term is three-dimensional and slowly varying, which means that is gives information about very short-range correlations within the ordered layers. This is most likely thermal diffuse scattering, from correlated atomic motion, as the measurement was carried out at 300 K or it could also be related to the slight Cu deficiency of the structure. The result of this is also seen in 3D-ΔPDF for the z = 0 layer close to the origin (Figure 4), where there are a few weak features not reproduced in the model 3D-ΔPDF. Further analysis of this very weak secondary diffuse term as a function of temperature might give insight into the mechanism of the Cu diffusion in the material.

### 5. Local coordination and dimer formation

Figure 6 shows two of the ordered layers stacked on top of each other where the different crystallographic sites are color-coded. Here the same mirror image of the layer is used with an interlayer vector of [-1/3, 0, 6.81Å], but all the different stacking types will give the same local coordinations. The site labels used here are the ones given in table 2.

The structure of $Cu_{2-x}Se$ consists of a rhombohedrally distorted face-centered cubic packing of Se with an alternating short (3.04Å) and long (3.78Å) distance between the close-packed layers. Figure S6 c and d show the coordination of the different Cu sites. The Cu1 sites (blue) are placed in tetrahedral holes in the short gap, pointing out of the ordered layer. The Cu2 (red), Cu3 (green) and Cu4 (orange) sites all have trigonal pyramidal coordination to Se, where Cu2 and Cu3 sites point into octahedral holes in the ordered layer, while the Cu4 sites point into tetrahedral holes. Cu2 sites form pairs pointing into the same octahedral hole from each side of the ordered layer while the Cu3 sites are adjacent to Cu1 sites. The Cu4 sites form dimers which share two Se atoms. The Cu4-Cu4 distance in these dimers is shorter (2.32Å) than the Cu4-Se distance (2.56Å). The formation of Cu-Cu dimers with short Cu-Cu distances down to 2.30 Å have previously been observed in molecular Cu-Se clusters (Dehnen *et al.*, 2002). Some of the largest of the reported molecular clusters with a 2:1 ratio of Cu:Se, e.g. $Cu_{140}Se_{70}(PEt_3)_{34}$, have three layers of Se stacked ABA (as in the hcp structure) with Cu in trigonal and tetrahedral coordination to Se. This is very similar to the structure for bulk $Cu_2Se$ found here, although the bulk structure has the fcc stacking of Se layers. These molecular clusters might be seen as precursor phases to the bulk structure and could give insight into the formation of $Cu_2Se$.

Half of one of the ordered layers, as viewed from above, is shown in figure 6 b. As seen in the figure, the Cu4 sites are displaced away from the ideal tetrahedral sites toward the Cu1 sites. As the Cu1 site is above the Se layer while the Cu2, Cu3 and Cu4 sites are below, there is more room underneath the Cu1 sites than the Cu2 and Cu3 sites.

### 6. Comparison with suggested structures in literature

A number of structures for β-$Cu_{2-x}Se$ have been suggested in the literature. However, none of them are in good agreement with the measured scattering data. Common to all of the proposed models is the assumption of a periodic structure in three dimensions in contrast to the real structure reported here, which is ordered in two dimensions and has stacking disorder in the third dimension. However, many of the suggested structures are periodic repetitions of structural units, which also occur in the real disordered structure. Gulay et al. suggested a monoclinic structure (spacegroup C2/c) with 4 layers going through the unit cell (Gulay *et al.*, 2011). The layers themselves are in fact equivalent to the 2D layer shown in Figure 5, but the stacking sequence is given by two identical layers, then two mirrored layers and so on. Liu *et al*. suggested two different structures based on DFT calculations (Liu, Yuan, *et al.*, 2013) with one being a monoclinic structure (spacegroup C2/c) and the other being triclinic (P-1). Both of these also have approximately the same ordered 2D layers as shown in Figure 5. The triclinic structure has a stacking of identical layers, whereas the monoclinic structure has a stacking of alternating mirrored and non-mirrored layers. It was noted that these two structures were very close in energy. Lu *et al.* suggested another triclinic structure (spacegroup P1), which has stacking of only one slightly distorted layer. They suggested that several of the proposed structures form together based on electron microscopy (Lu *et al.*, 2015). Qiu et al. suggested three additional structures with very small energy differences, and based on electron microscopy measurements they also suggested that the structures coexist.

They suggested two triclinic structures (spacegroup P-1) and one monoclinic structure (spacegroup C2/c), which all have slightly distorted layers compared to the ones found here (Qiu *et al.*, 2016). Both the triclinic structures have stacking of one layer type (no mirrored layers) and the monoclinic structure has alternating mirrored and non-mirrored layers. Nguyen et al. suggested a monoclinic structure (P21/c), which does not resemble the real structure reported here (Nguyen *et al.*, 2013). Eikeland *et al.* proposed a monoclinic superstructure (C2/c) with two types of layers stacking in an alternating sequence. One of the layer types is the correct ordered layer while the other does not resemble the correct ordered layer (Eikeland *et al.*, 2017).

As several of the suggested structures found in the literature which have the correct layer types, but periodic stacking sequences, have been noted to have very similar enthalpies of formation, the stacking disorder found in the real structure is caused by the almost identical enthalpy of the different stacking types together with a higher entropy term from the disorder.

## 7. Concluding remarks

High-energy synchrotron radiation in combination with low noise detectors has made it possible to measure weak diffuse scattering signals from advanced materials in large volumes of reciprocal space. From such data it is becoming increasingly evident that structural disorder is present e.g. in key thermoelectric materials such as PbTe and $Cu_{2-x}Se$, and, moreover, that this disorder is critical for giving the materials their desirable properties. The apparent disorder is not fully random, but correlated, and by using 3D-ΔPDF analysis of single-crystal diffuse X-ray scattering we show that it is possible to solve structures that are ordered in some dimensions but disordered in others. Correlated disorder means that there is a local order in regions of the crystal dictated by specific chemical bonding interactions. The term local order might therefore be more describing than correlated disorder.

Here we have studied the thermoelectric material β-$Cu_{2-x}Se$ for which the structure has been discussed since 1936 without any satisfying structural models being able to explain the measured scattering data (Eikeland *et al.*, 2017; Frangis *et al.*, 1991; Gulay *et al.*, 2011; Kashida & Akai, 1988; Liu, Yuan, *et al.*, 2013; Lu *et al.*, 2015; Milat *et al.*, 1987; Nguyen *et al.*, 2013; Qiu *et al.*, 2016; Rahlfs, 1936; Yamamoto & Kashida, 1991). The reason for this failure has been that all previous structural models have assumed three-dimensional periodicity as well as difficulties in synthesizing high quality single crystals. The real structure of β-$Cu_{2-x}Se$ consists of two-dimensional ordered layers, which stack in a highly disordered sequence. Using the 3D-ΔPDF, the structure of the ordered 2D layer could be determined by methods related to crystallographic Patterson analysis. In addition, the precise modes of stacking disorder could be identified. Two disorder modes are present in the stacking, one being the sequence of mirrored and non-mirrored layers and the other being the sequence of three possible inter-layer vectors. The disordered stacking gives rise to an average structure with higher symmetry. Several of the previously suggested structures in the literature, which have been noted to have very small differences in enthalpy, in fact have the correct 2D layer structure, but they all assume various periodic stacking sequences (Gulay *et al.*, 2011; Liu, Yuan, *et al.*, 2013; Lu *et al.*, 2015; Qiu *et al.*, 2016). The stacking disorder found in the real structure is caused by the almost identical enthalpy of the different stacking types together with a higher entropy term from the disorder. This underlines the importance of considering non-periodic structures in materials science. In general, the crystal structure induces the specific physical properties of a material, and any kind of quantitative modelling of real materials therefore must use the correct crystal structure. We have constructed the new structural model for $Cu_{2-x}Se$ based on a combination of standard crystallographic analysis of Bragg diffraction data, giving the average structure, and 3D-ΔPDF analysis of the diffuse scattering, giving the deviations from the average structure. The 3D-ΔPDF gives an intuitive and direct method for structural solution of disordered systems without the need for reverse Monte-Carlo simulation or energy minimization calculations.

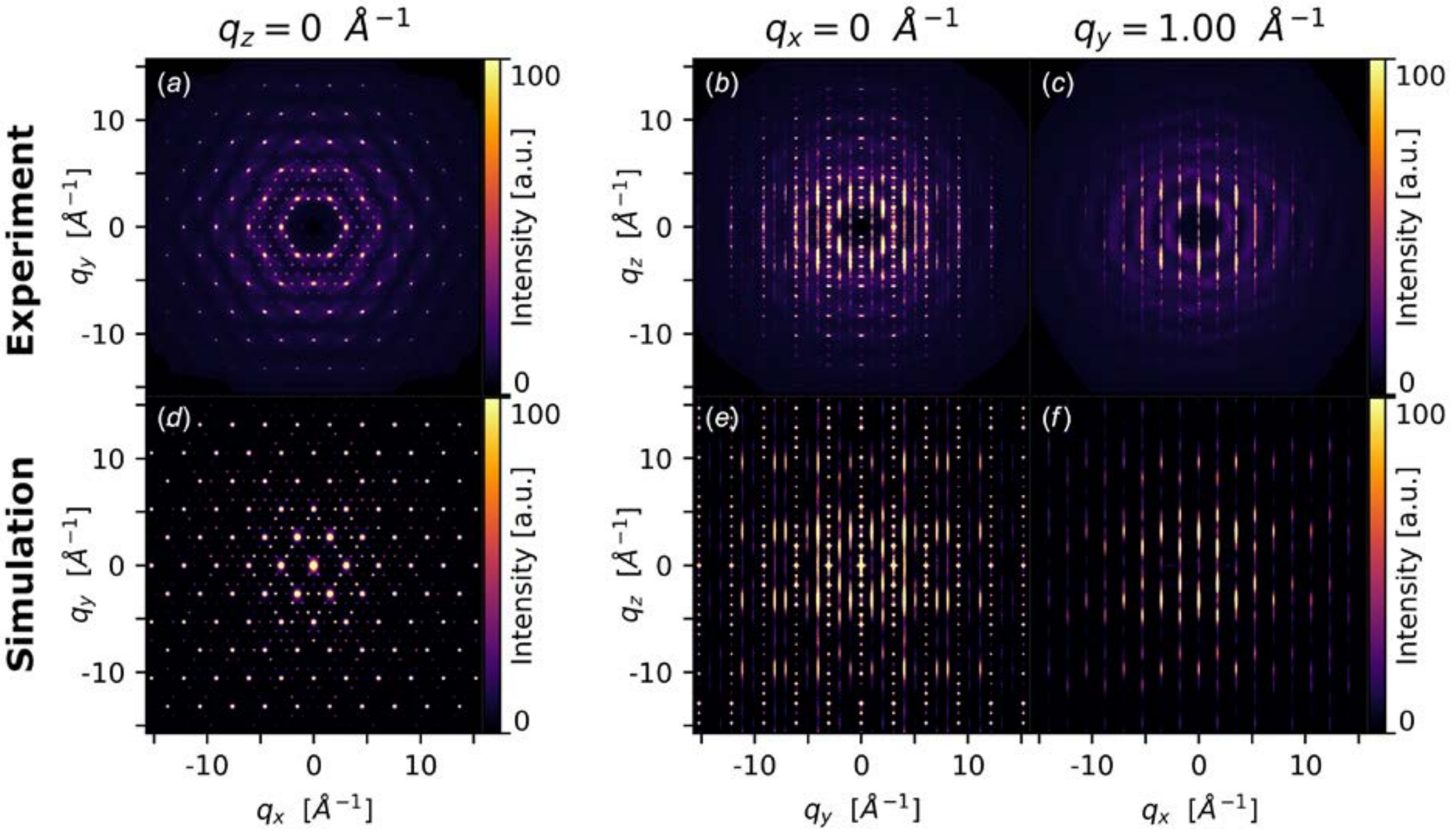

**Figure 1** Experimental (a,b,c) and calculated (e,d,f) X-ray scattering for β-$Cu_{1.95}Se$ at 300 K. (a) $q_z = 0$ plane with strong reflections which give an average structure with space group R$\bar{3}$m, while weak reflections are seen between the main reflections. (b) $q_x = 0$ plane where rows of Bragg peaks are seen with lines of diffuse scattering between them. (c) $q_y = 1.00$ Å$^{-1}$ plane which only has diffuse scattering lines. (d), (e) and (f) are the corresponding layers of the calculated scattering. The calculation is based on a random stacking of layers generated with both a random order of mirrored and non-mirrored layers as well as a random order of the three possible inter-layer vectors (see text). The calculation does not take atomic vibrations into account.

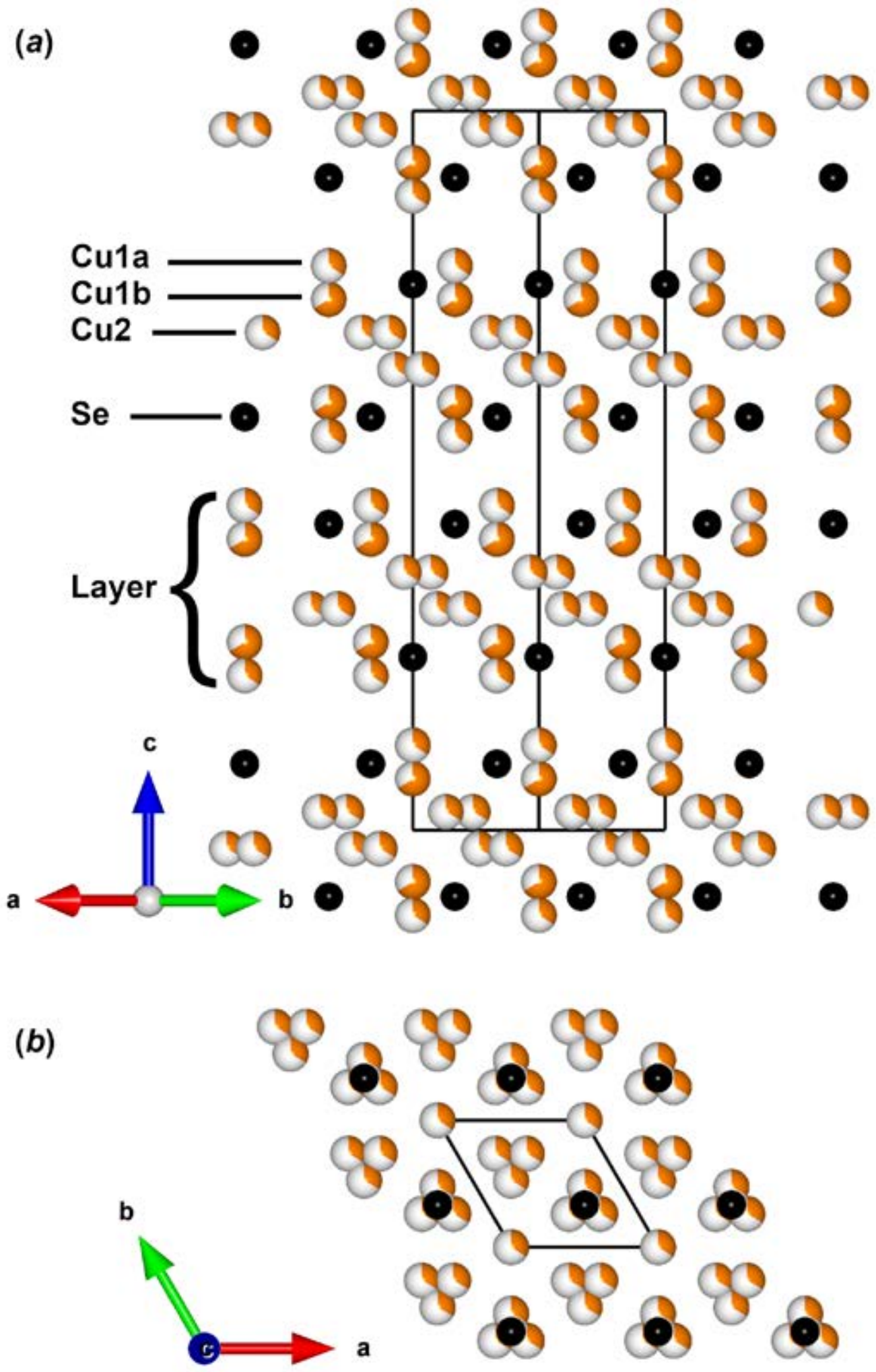

**Figure 2** Average structure of β-$Cu_{1.95}Se$. **a**: The layered structure with the different sites marked. **b**: A single layer seen from above. The black atoms are Se, and the orange are Cu. The partly filled colors show the occupancy of the site.

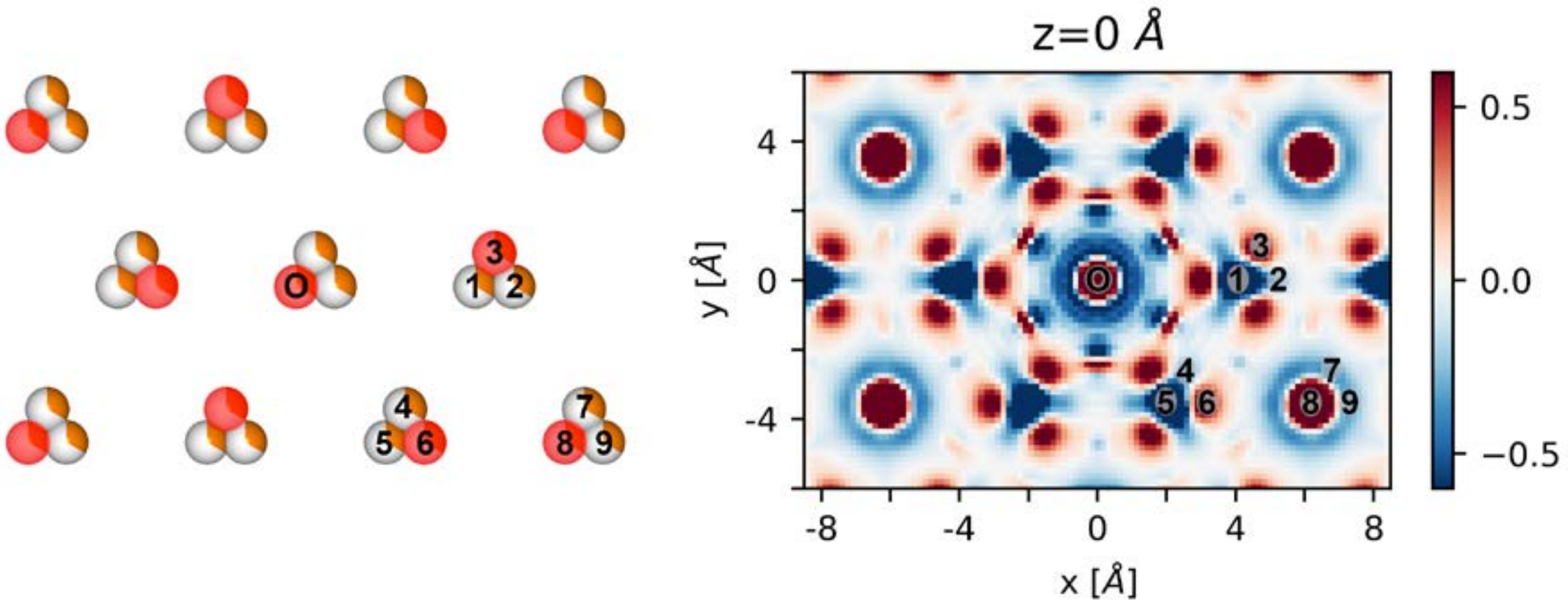

**Figure 3** Two-dimensional ordering of the Cu2 layers in β-$Cu_{1.95}Se$. (Left) A layer of Cu2 sites in the average structure, which form triangles with each triangle 1/3 occupied. Red-tinted atoms mark the ordered structure if a Cu atom occupies the position marked “O”. (Right) The z = 0 plane of the 3D-ΔPDF for short range correlations. Numbers, n, marked in the left figure give vectors O-n which are marked by the same number in the right figure.

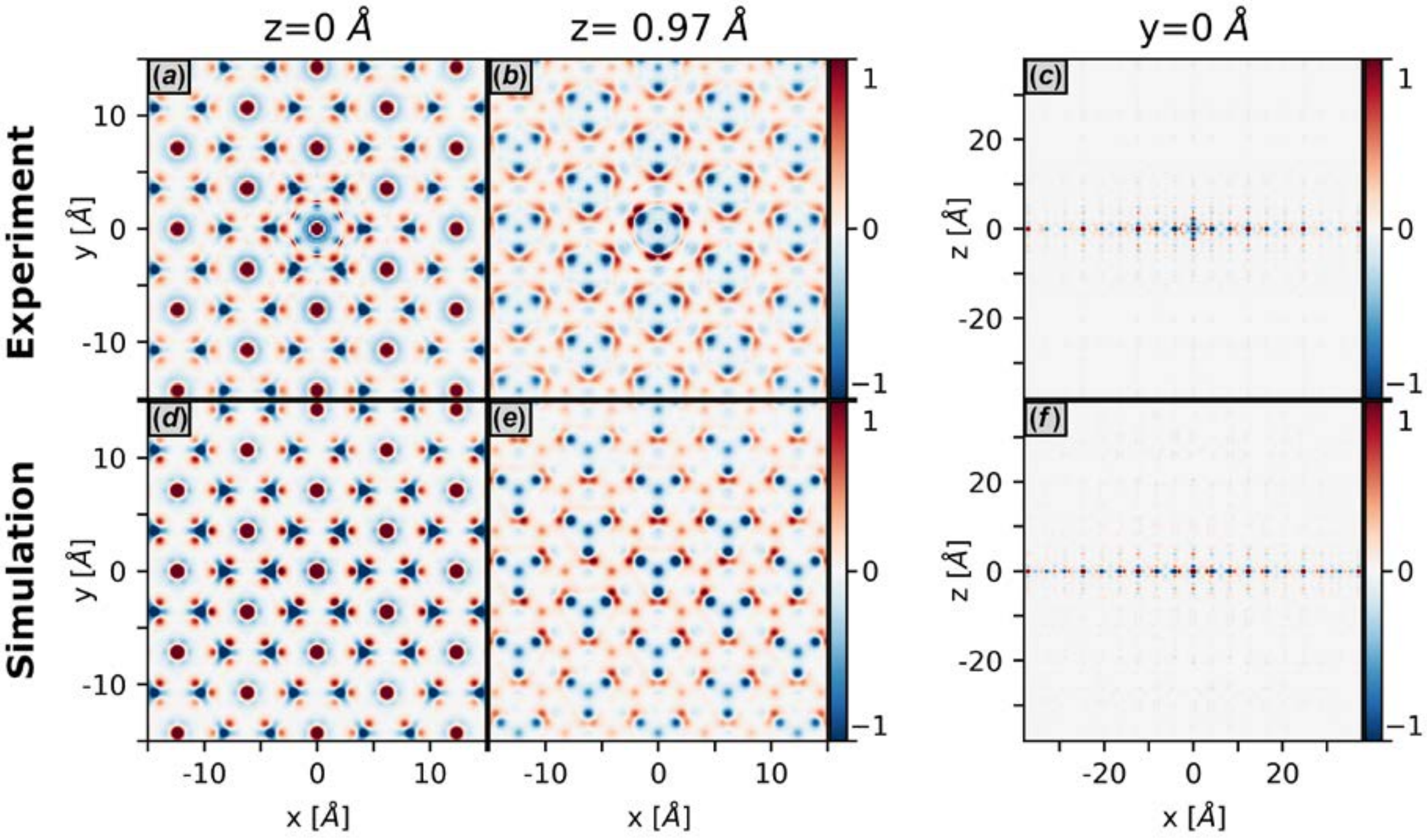

**Figure 4** 3D-ΔPDF for $Cu_{2-x}Se$ at 300K. (a, b, c) are the measured 3D-ΔPDF while (d,e,f) are the corresponding calculated 3D-ΔPDF from the simulated model of perfect two-dimensional order and one-dimensional disorder. The simulated 3D-ΔPDF has been broadened by a Gaussian function to emulate the effect of isotropic thermal vibration.

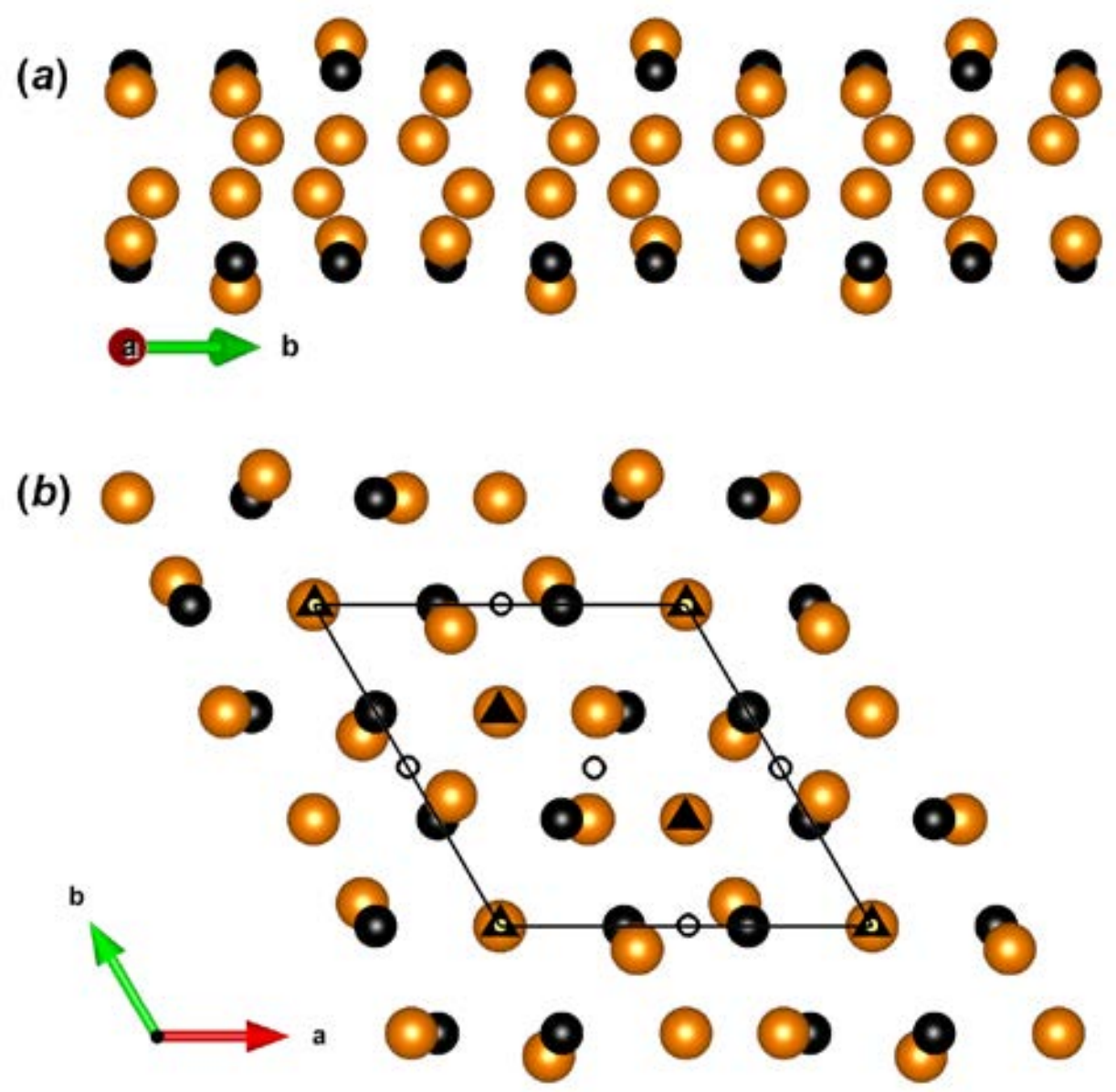

**Figure 5** 2D layer superstructure of β-$Cu_{1.95}Se$. (a) Layer seen from the side. (b) A single layer seen from above with the layer group symmetry marked. The black atoms are Se, and the orange are Cu. A triangle marks a three-fold axis; a circle marks an inversion center.

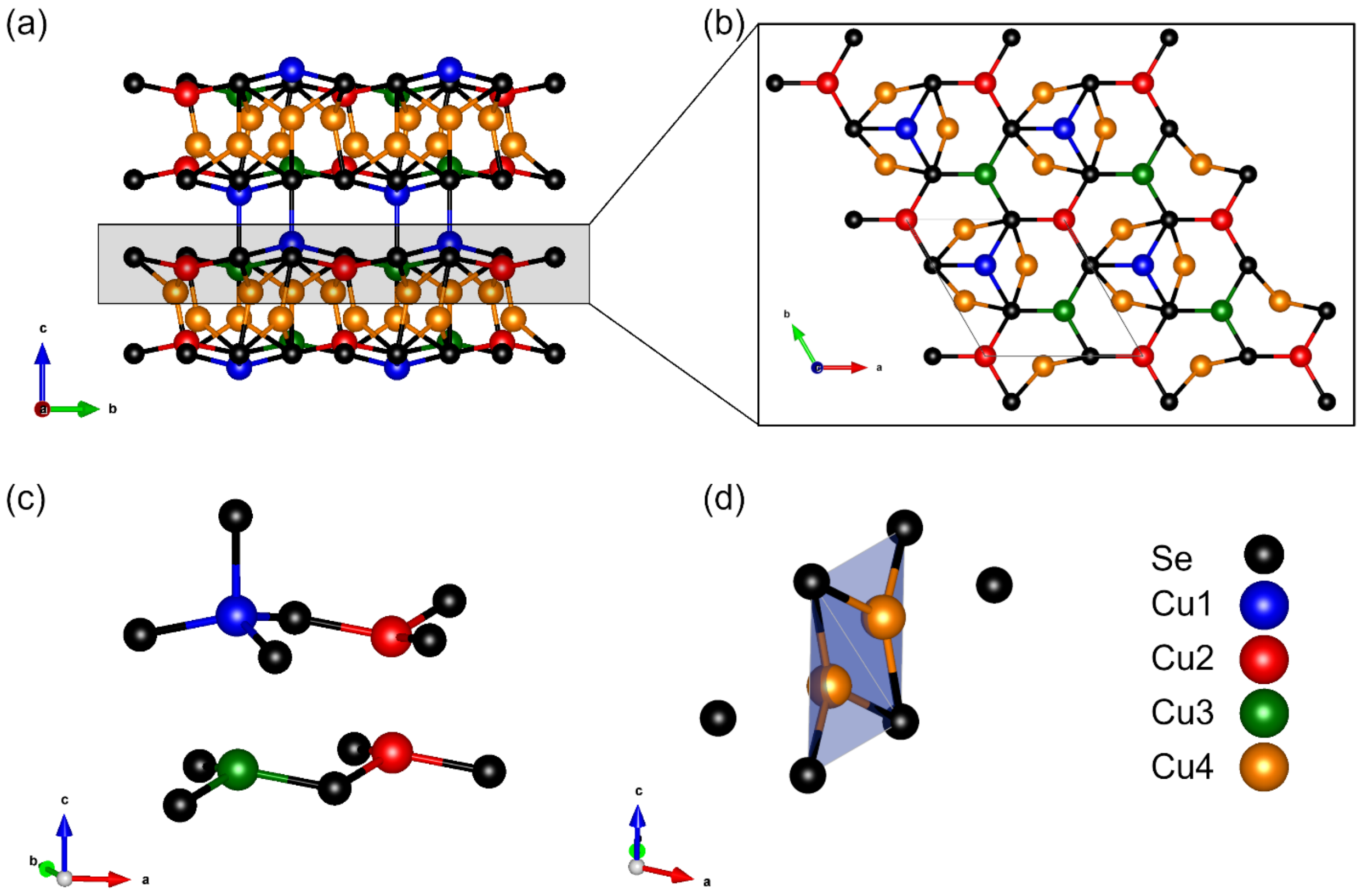

**Figure 6** Local coordination in $Cu_{2-x}Se$ with color-coding of different sites. (a) Two layers stacked on top of each other, here using the same mirror image and the inter-layer vector [-1/3,0,0.681Å]. The gray box marks half of one layer, which is shown in (b). (b) A close-packed Se layer together with close-contact Cu. (c) Trigonal pyramidal coordination of Cu1, Cu2 and C3 sites. (d) Trigonal pyramidal coordination of two Cu4 sites forming a dimer. The site labels used here refer to table 2.

**Table 1** Crystallographic data for the average structure

| | |
|---|---|
| Chemical Formula | $Cu_{1.95(2)}Se$ |
| Space group | $R\bar{3}m$; H |
| Temperature | 300K |
| *a* (Å) | 4.1217(3) |
| *c* (Å) | 20.435(3) |
| *V* (Å$^3$) | 300.65(7) |
| Z | 6 |
| $\rho_{calc}$ (g cm$^{-3}$) | 6.726 |
| $\mu$ (mm$^{-1}$) | 38.35 |
| F(000) | 544 |
| $(\sin\theta / \lambda)_{max}$ (Å$^{-1}$) | 0.74 |
| $N_{Tot,obs}$ | 2127 |
| $N_{Uniq,obs}$ | 157 |
| $N_{Parameters}$ | 19 |
| GOF | 1.107 |
| $R_{int}$ | 0.0797 |
| $R_1$ | 0.0404 |
| $wR_2$ | 0.0804 |

| | Position [x/a, y/b, z/c] | Wyckoff position | Occupancy |
|---|---|---|---|
| Se | [0, 0, 0.24097 (7)] | 6c | 1 |
| Cu1a | [0, 0, 0.1181 (6)] | 6c | 0.328(14) |
| Cu1b | [0, 0, 0.0715 (4)] | 6c | 0.648(14) |
| Cu2 | [0.2624 (6), 0.5248(6), 0.02557(16)] | 18h | 0.325(5) |

**Table 2** Structure of ordered layer in β-$Cu_{1.95}Se$. Note that x and y coordinates are given relative to the unit cell axis, while the z coordinate is in Å. The mirrored layer with the operation (x,y,z)→(y,x,z) is also permitted.

| Layer group | | | | $p\bar{3}$ (#66) | |
|---|---|---|---|---|---|
| Temperature | | | | 300K | |
| *a* (Å) | | | | 7.1389(5) | |
| | x/a | y/b | z | Wyckoff position | Corresponding site in average structure |
| Se | 1/3 | 1/3 | 1.888Å | 6e | Se |
| Cu1 | 1/3 | 2/3 | 2.415Å | 2c | Cu1a |
| Cu2 | 0 | 0 | 1.463Å | 2b | Cu1b |
| Cu3 | 2/3 | 1/3 | 1.463Å | 2c | Cu1b |
| Cu4 | 0.596 | 2/3 | 0.523Å | 6e | Cu2 |

**Acknowledgements** The work was supported by the Danish National Research Foundation (DNRF93). The authors gratefully acknowledge Martin von Zimmermann for sharing his script for converting raw data to reciprocal space. Kristoffer Holm, Lennard Krause and Emil Klahn are gratefully acknowledged for help with data measurements. Espen Eikeland and Kasper Tolborg are thanked for fruitful discussions. NSF's ChemMatCARS Sector 15 is principally supported by the Divisions of Chemistry (CHE) and Materials Research (DMR), National Science Foundation, under grant number NSF/CHE-1346572. Use of the Advanced Photon Source, an Office of Science User Facility operated for the U.S. Department of Energy (DOE) Office of Science by Argonne National Laboratory, was supported by the U.S. DOE under Contract No. DE-AC02-06CH11357.

# Supporting information

## S1. Interpretation of the 3D-ΔPDF peak amplitudes

Starting from the equation given for the 3D-ΔPDF given in the introduction and partitioning the electron density into a sum of atomic electron densities, the 3D-ΔPDF can be rewritten as

$$\text{3D-}\Delta\text{PDF}(\boldsymbol{r}) = \sum\nolimits_{ij} \delta\rho_j(\boldsymbol{r}) * \delta\rho_i(\boldsymbol{r}) * \delta(\boldsymbol{r} - \boldsymbol{r}_{ij})$$

Where $\delta\rho_i(\boldsymbol{r})$ is the difference between the real and average electron density of the i'th atom, $\delta(\boldsymbol{r})$ is the Dirac-delta function, $\boldsymbol{r}_{ij}$ is the vector between atom i and atom j, and $*$ is the convolution operator. For every pair of atoms there will be a peak in the 3D-ΔPDF given by $\delta\rho_j(\boldsymbol{r}) * \delta\rho_i(\boldsymbol{r})$ at the position $\boldsymbol{r}_{ij}$. Because of the Fubini-Tonelli theorem, the integral of such a peak will be given by the product of the integral of the two functions, and therefore just as the product of the difference in number of electrons:

$$\text{peak amplitude} \propto \int_{peak} \text{3D-}\Delta\text{PDF}\,(\boldsymbol{r})\, d\boldsymbol{r} = \int \delta\rho_j(\boldsymbol{r})\, d\boldsymbol{r} \cdot \int \delta\rho_i(\boldsymbol{r})\, d\boldsymbol{r} = \delta Z_j \cdot \delta Z_i$$

where $\delta Z_i$ is the difference in number of electrons in the real and average structure for atom i (e.g. if the atom is occupied 1/3 in the average structure, $\delta Z$ will be $\frac{+2}{3}Z$ if the site is occupied and $\frac{-1}{3}Z$ if the site is unoccupied, where Z is the atomic number).

## S2. Solving the structure of β-$Cu_{2-x}Se$ "by hand"

The ordered structure needs to have a periodic structure in the plane with unit cell vectors $a_{layer} = a_{avg} - b_{avg}$ and $b_{layer} = 2b_{avg} + a_{avg}$, as this gives rise to the positions of the sharp "superstructure" lines in the scattering, as shown in Figure S1. The same can be seen in the 3D-ΔPDF which is periodic in two dimensions. The unit cell of the average structure is marked by a green box and the ordered unit cell as a black box in the right part of Figure S1. It is seen that the average unit cell vectors have negative peaks in the 3D-ΔPDF while the ordered unit cell vectors are at positive peaks, validating the dimensions of the ordered unit cell.

Each layer in the average structure contains two sub-layers of each type (Se, Cu1a, Cu1b, Cu2). In Figure S2 one such layer is shown with marked sub-layers for reference.

First the individual layers are evaluated. Each of the three types of Cu layers together with the 3D-ΔPDF for z = 0 Å are shown in Figure S3. To find the ordered structure for each layer it is first assumed that the atom marked "O" is occupied (The choice of a different "O" would give the same structure, just with a shifted origin). The vectors to atoms with marked numbers are marked by the same numbers in the 3D-ΔPDF. The solution for the layers are shown with red-tinted atoms. The solution for the Cu2 layer is given in the main text. The Cu1a layer is 1/3 occupied and the Cu1b layer 2/3 occupied. For each of these layers there is only one possible solution in agreement with the ordered unit cell size. The solution to the Cu1a layer is also clearly in agreement with the 3D-ΔPDF, where the vectors between occupied sites have positive peaks while the vectors from occupied to non-occupied have negative peaks. The agreement of the Cu1b layer with the 3D-ΔPDF is less clear. We look at vector to the nearest neighbor (e.g. O-1). When going through the sites, one third of the time this vector separates two occupied sites (e.g. 5 and 8) while two thirds of the time the vector separates an occupied and an unoccupied site (e.g. O and 1). This gives a peak amplitude proportional to $\delta Z_{occ}^2 + 2\,\delta Z_{occ}\,\delta Z_{unocc} = \left(\left(\frac{1}{3}\right)^2 + 2\frac{1}{3}\frac{-2}{3}\right) Z_{Cu}^2 = \frac{-1}{3} Z_{Cu}^2$ .

This is in agreement with the observed negative feature for vector 1.

Now that the ordered structures for all layer types has been identified, the inter-layer orderings can be found. Only three inter-layer orderings are needed to construct the full layer: Cu1a to Cu1b, Cu1b to Cu2 and Cu2 to Cu2. Table S1 gives the vertical distances between all different sub-layers within the layer, where the notation for the sublayers is given in Figure S2. The ordering of these sublayers to each other can then be found from the 3D-ΔPDF for the vertical distance between the layers. Figure S4 shows one such layer for z=0.97Å.

First we look at the ordering between the close Cu1a and Cu1b layers (e.g. Cu1a_1 and Cu1b_a in Figure S2). As the Cu1a and Cu1b sites are only 0.95 Å apart, it is not possible for a Cu1a site right next to a Cu1b site to both be filled at the same time. As 1/3 of the Cu1a sites and 2/3 of the Cu1b sites are filled, either the Cu1a or the Cu1b site is filled for every pair of sites. As we know how each of these layers order, there is only one possibility to combine the two, which is illustrates in Figure S5. The same is seen in the 3D-ΔPDF for the z = 97Å layer in Figure S5 (with the grid size used here, 0.97Å is the closest layer of the obtained 3D-ΔPDF to 0.95Å). For the vector (x,y,z) = (0, 0, 0.97Å) there is a negative feature, showing there to be no Cu1a and Cu1b sites filled right on top of each other as expected. For the vectors corresponding to the six surrounding atoms in the other layer, positive features are observed, showing the Cu1a site to be surrounded by filled Cu1b sites. For the vectors corresponding to the ordered unit cell in the xy-plane, there are negative features in the 3D-ΔPDF for this z-coordinate, again showing that the Cu1a and Cu1b sites right on top of each-other are not filled at the same time.

We then look at the coupling from Cu1b to Cu2 layers next to each other (e.g. Cu1b_1 and Cu2_1 in Figure S2). These have a vertical distance of 0.94 Å. These layers are shown in Figure S6. As the ordering within these two layers are known, again only the relative positon of the layers is needed. First the layer of Cu1b site is filled with the known ordered pattern, shown by the blue-tinted atoms. There are two types of occupied Cu1b sites, marked O and O' and one unoccupied type of Cu1b site, marked O*. The vectors from O to site 3, 4 and 5 are identical to the vectors from O' to 3', 4' and 5' and O* to 3*,4* and 5*. The 3D-ΔPDF for the vector 3 is negative, showing that site 3 and 3' are not occupied. If just one of 3 or 3' were occupied, the 3D-ΔPDF would have been positive. If e.g. 3 was occupied, the O-3, O’-3’ and O*-3* would contribute to the 3D-ΔPDF as $\delta Z_O \delta Z_3 + \delta Z_{O'} \delta Z_{3'} + \delta Z_{O^*} \delta Z_{3^*}$

$$= \left(1 - \frac{2}{3}\right)\left(1 - \frac{1}{3}\right) Z_{Cu}^2 + \left(1 - \frac{2}{3}\right)\left(0 - \frac{1}{3}\right) Z_{Cu}^2 + \left(0 - \frac{2}{3}\right)\left(0 - \frac{1}{3}\right) Z_{Cu}^2 = \frac{1}{3} Z_{Cu}^2$$

As the ordered structure of the Cu2 layer is known, there is only one possible combination of the two layers that gives 3 and 3’ not occupied. This ordering is marked by the red tinted atoms in Figure S6. The vectors 4 and 5 in the 3D-ΔPDF have positive features, in agreement with the O-4 and O’-5’ occupied sites.

Finally we look for the relation between two Cu2 layers (e.g. Cu2_1 and Cu2_2 in Figure S2). Figure S7 show these layers. It is first assumed that the lower layer has the blue-tinted sites occupied. We look at the occupied site O and the two unoccupied sites O’ and O*. We then look at the vector O-6 which is identical to O’-6’ and O*-6*. The 3D-ΔPDF has a negative feature for this vector, which is only possible if there are no atoms separated by that vector, meaning that site 6 has to be empty. If site 6 was occupied the 3D-ΔPDF for this vector would be positive as $\delta Z_O \delta Z_6 + \delta Z_{O'} \delta Z_{6'} + \delta Z_{O^*} \delta Z_{6^*}$

$$= \left(1 - \frac{1}{3}\right)\left(1 - \frac{1}{3}\right) Z_{Cu}^2 + \left(0 - \frac{1}{3}\right)\left(0 - \frac{1}{3}\right) Z_{Cu}^2 + \left(0 - \frac{1}{3}\right)\left(0 - \frac{1}{3}\right) Z_{Cu}^2 = \frac{2}{3} Z_{Cu}^2$$

There are two possible arrangements that leave site 6 empty. These are shown in Figure S7 by the red-tinted atoms. In one of the possibilities site 7 is empty while site 8 is occupied and vice versa in the other possibility. In the 3D-ΔPDF it is seen that vectors 7 and 8 both have positive peaks. There is a negative peak for vector 9, showing that site 9 needs to be empty. Positive peaks are observed for vectors 10 and 11, where one of the possible arrangements has 10 occupied and the other has 11 occupied. This shows that both possibilities are present in the real structure. The two possibilities are the mirror image of each other. This is where one of the modes of disorder in the structure is introduced.

The full layer can now be constructed as each sublayer is known as well as the ordering between these. Two types of layers are needed, one for each of the possible orderings between two Cu2 layers, as shown in Figure S7. These two types of layers are related through a mirror plane.

### S3. Simulations for different models.

Simulations were also made for models where only of the two modes of disorder (mirror/non-mirror layers and inter-layer vectors) were used.

If the simulation is made only using the disorder in the stacking vectors and all layers are the same mirror image, 3D-ΔPDF it not in agreement with experiment. The simulated 3D-ΔPDF for the z = 0.97Å layer for this model is shown in Figure S8 and should be compared to the experiment in Figure S4.

Likewise if only the disorder in whether each layer is mirrored or not is used, the 3D-ΔPDF is also not in agreement with experiment. For this model the 3D-ΔPDF for the y = 0Å layer is shown in Figure S9. In this case strong features are seen for every 20.4 Å, corresponding to 3 layers in the structure. This should be compared to the experimental 3D-ΔPDF as shown in Figure 4 in the main text, where the features quickly disappear along the *z* direction.

**Table S1** Distances between sub-layers in the ordered layer stack.

| Vertical distance [Å] | 1st layer | 2nd layer |
|---|---|---|
| 0.42 | Cu1b_1 | Se_1 |
| 0.53 | Cu1a_1 | Se_1 |
| 0.94 | Cu1b_1 | Cu2_1 |
| 0.95 | Cu1a_1 | Cu1b_1 |
| 1.05 | Cu2_1 | Cu2_2 |
| 1.36 | Cu2_1 | Se_1 |
| 1.89 | Cu1a_1 | Cu2_1 |
| 1.99 | Cu1b_1 | Cu2_2 |
| 2.41 | Cu2_1 | Se_2 |
| 2.93 | Cu1b_1 | Cu1b_2 |
| 2.94 | Cu1a_1 | Cu2_2 |
| 3.34 | Cu1b_1 | Se_2 |
| 3.88 | Cu1a_1 | Cu1b_2 |
| 4.30 | Cu1a_1 | Se_2 |
| 4.83 | Cu1a_1 | Cu1a_2 |

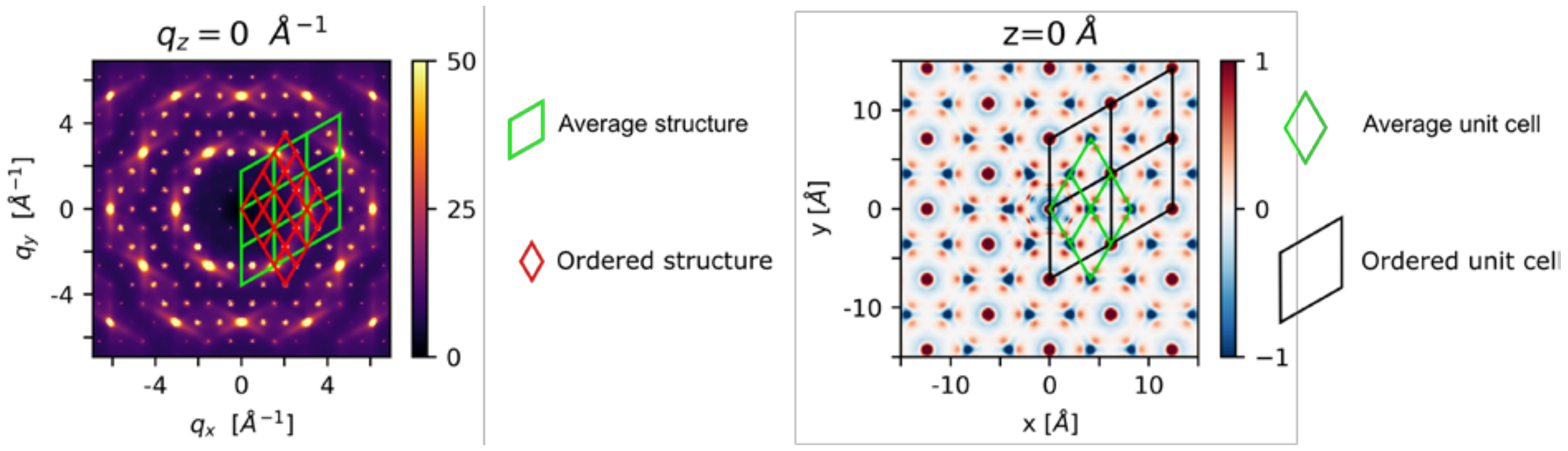

**Figure S1** Reciprocal and direct lattice of $Cu_2Se$ for the average and ordered structures. Left: $q_z = 0$ plane for the scattering, with the average reciprocal lattice in green and the ordered reciprocal lattice in red. Right: Z = 0 Å plane of the measured 3D-ΔPDF with the average unit cell in green and the ordered unit cell in black.

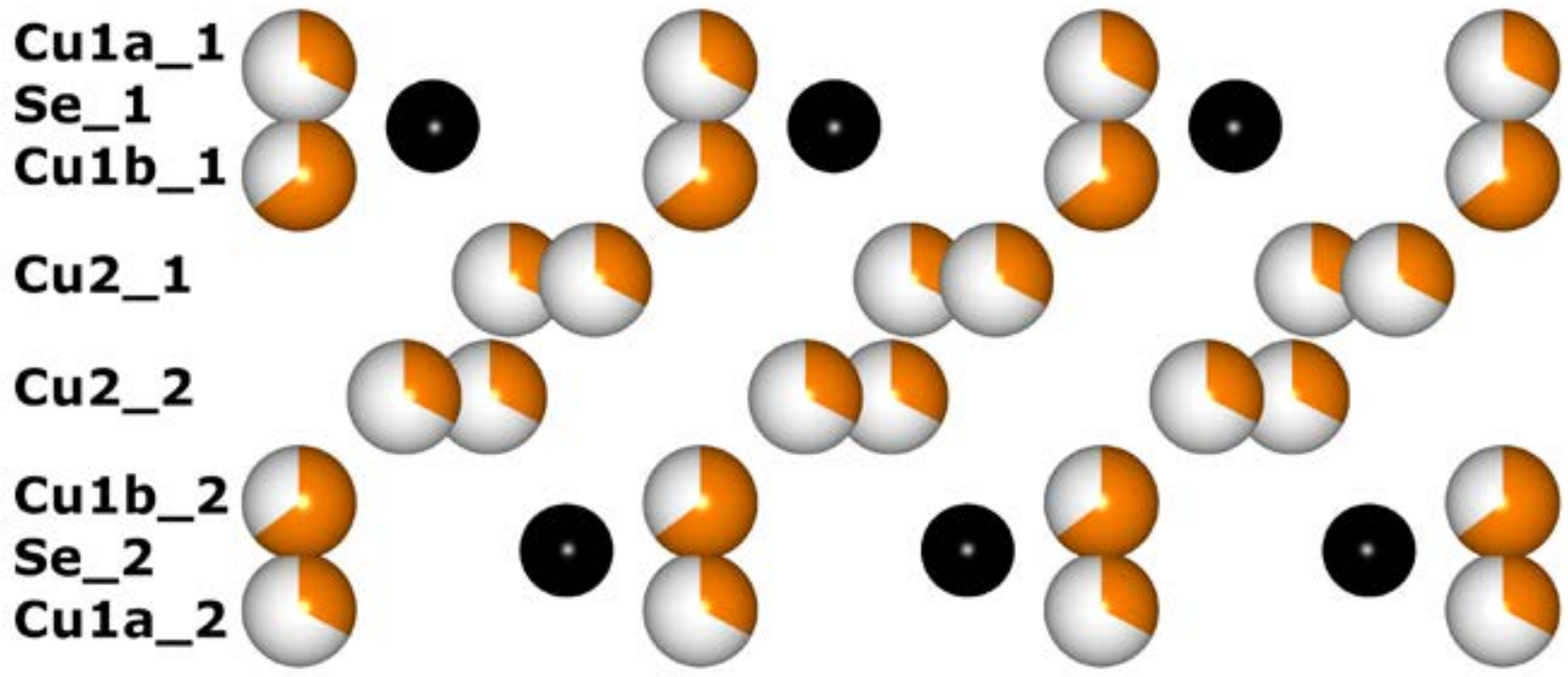

**Figure S2** Average layer structure. Partially filled atoms show the degree of occupancy. Black is Se and orange is Cu.

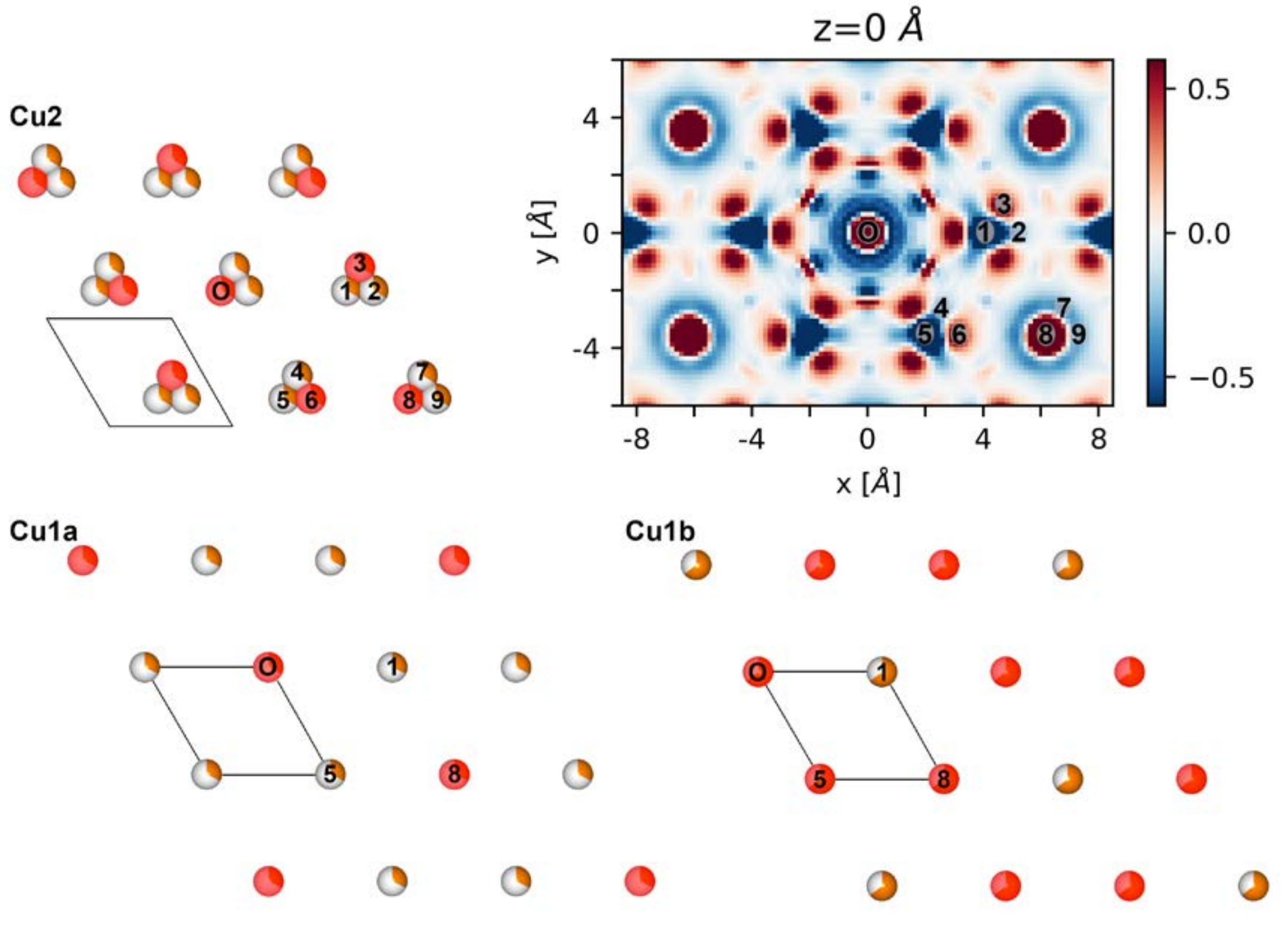

**Figure S3** Structure of Cu layers and the 3D-ΔPDF for the z=0Å plane. The ordered structure for each layer is marked with red-tinted circles. Vectors O-n for the numbers, n, marked in the structures are marked by the same numbers in the 3D-ΔPDF. The black lines mark the unit cell of the average structure

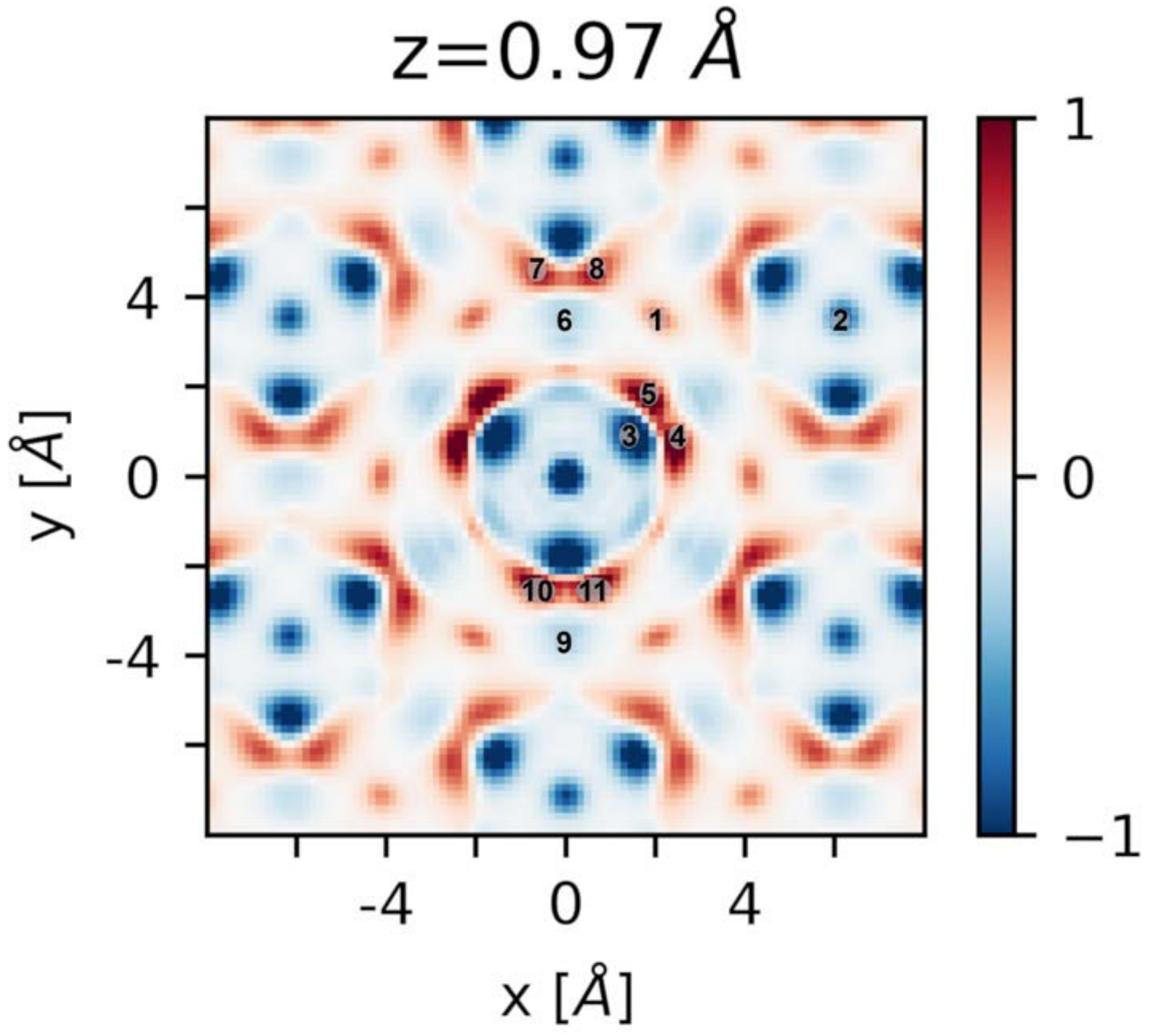

**Figure S4** 3D-ΔPDF for the z=0.97Å layer.

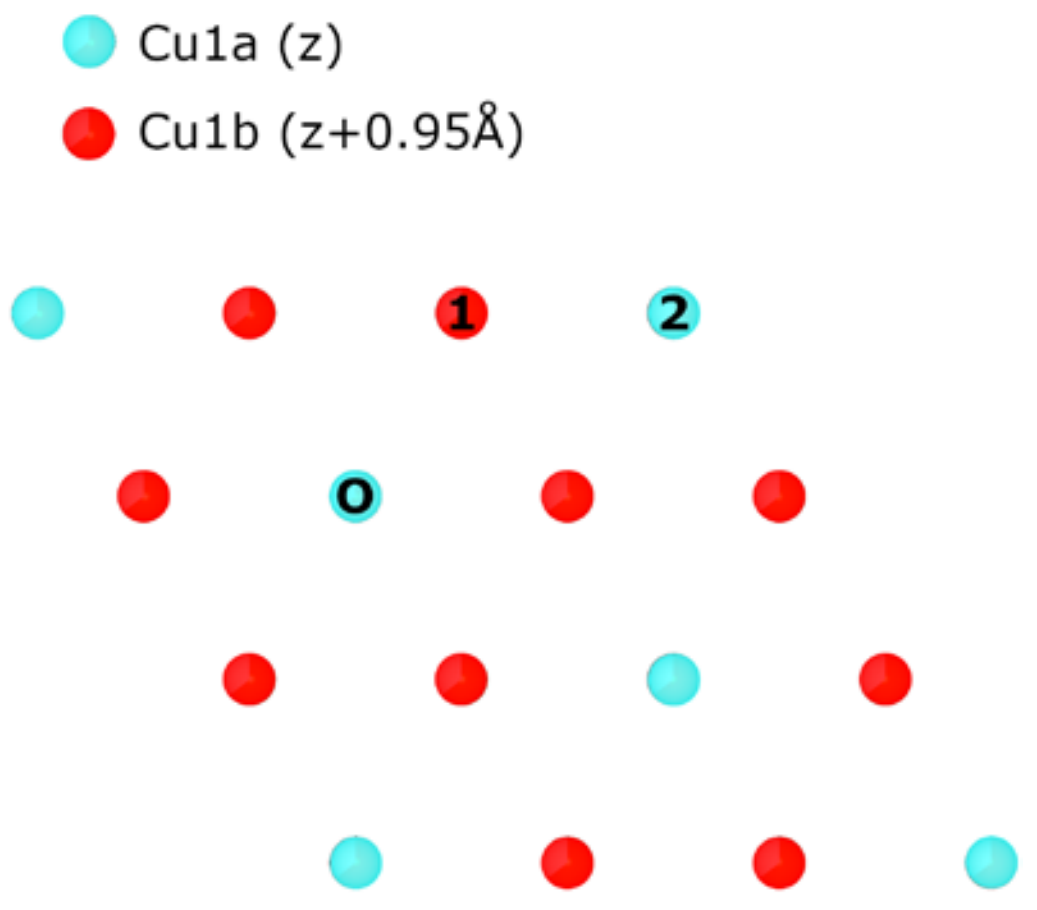

**Figure S5** Ordering of close-contact Cu1a and Cu1b layers. The Cu1a and Cu1b layer are separated in z by 0.95Å.

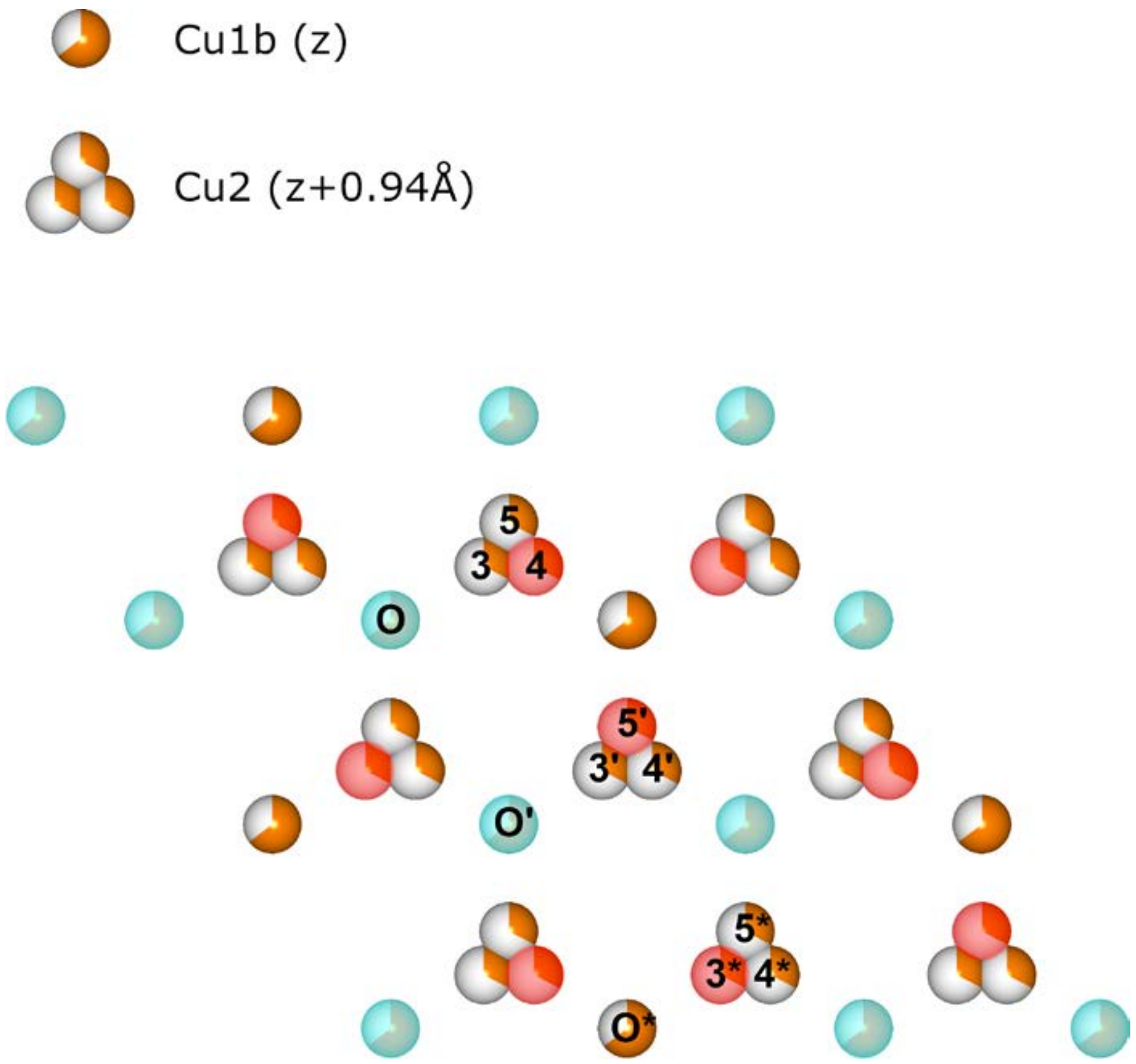

**Figure S6** Ordering of close Cu1b and Cu2 layers. The Cu1b and Cu2 layer are separated in z by 0.94Å. Blue and red tinted atoms mark the occupied sites in the Cu1b and Cu2 layer, respectively.

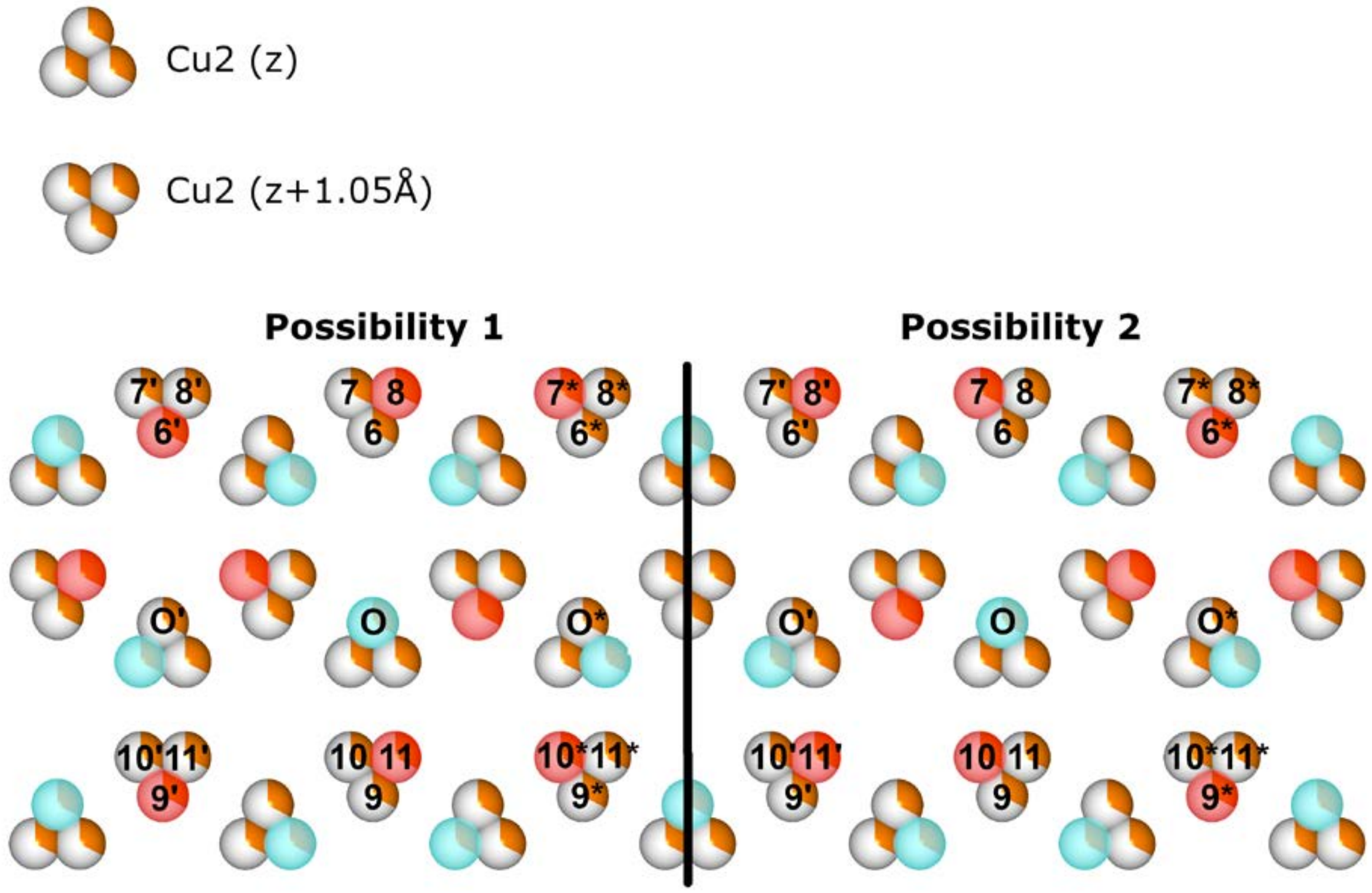

**Figure S7** Ordering of close Cu2 layers. The layers are separated in z by 1.05 Å. Blue and red tinted atoms mark the occupied sites in the lower and upper layer, respectively.

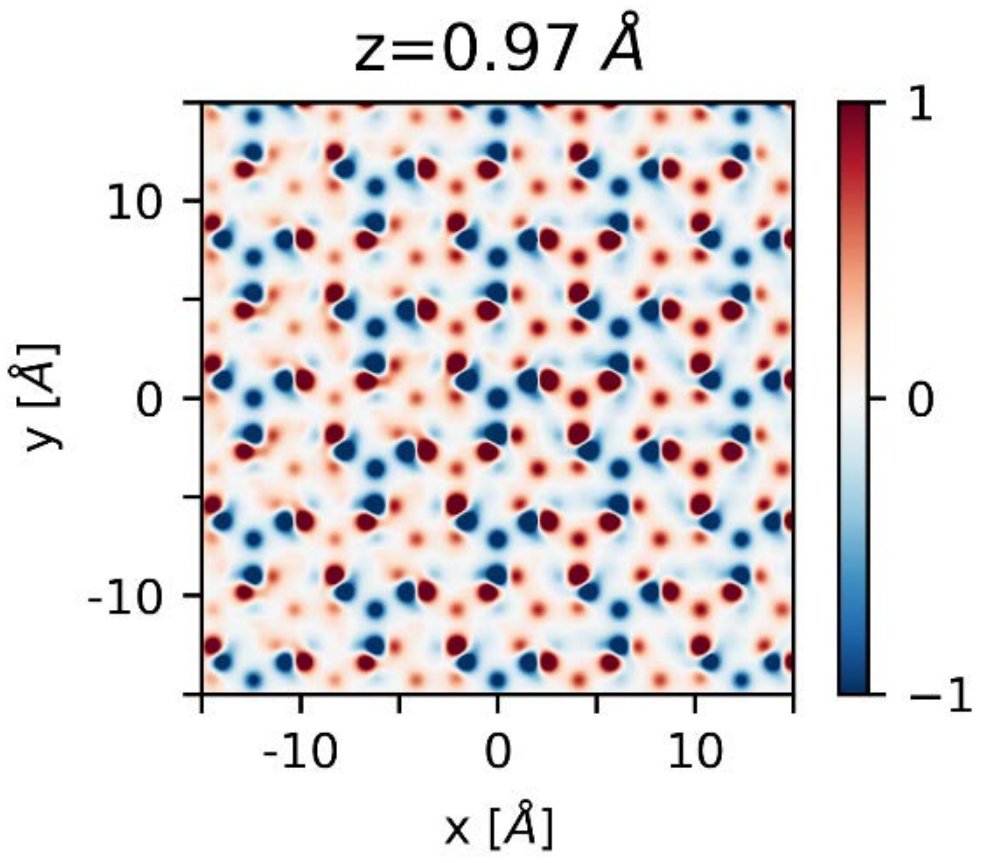

**Figure S8** Simulated 3D-ΔPDF for the model where the mirror image disorder of the layer is not used.

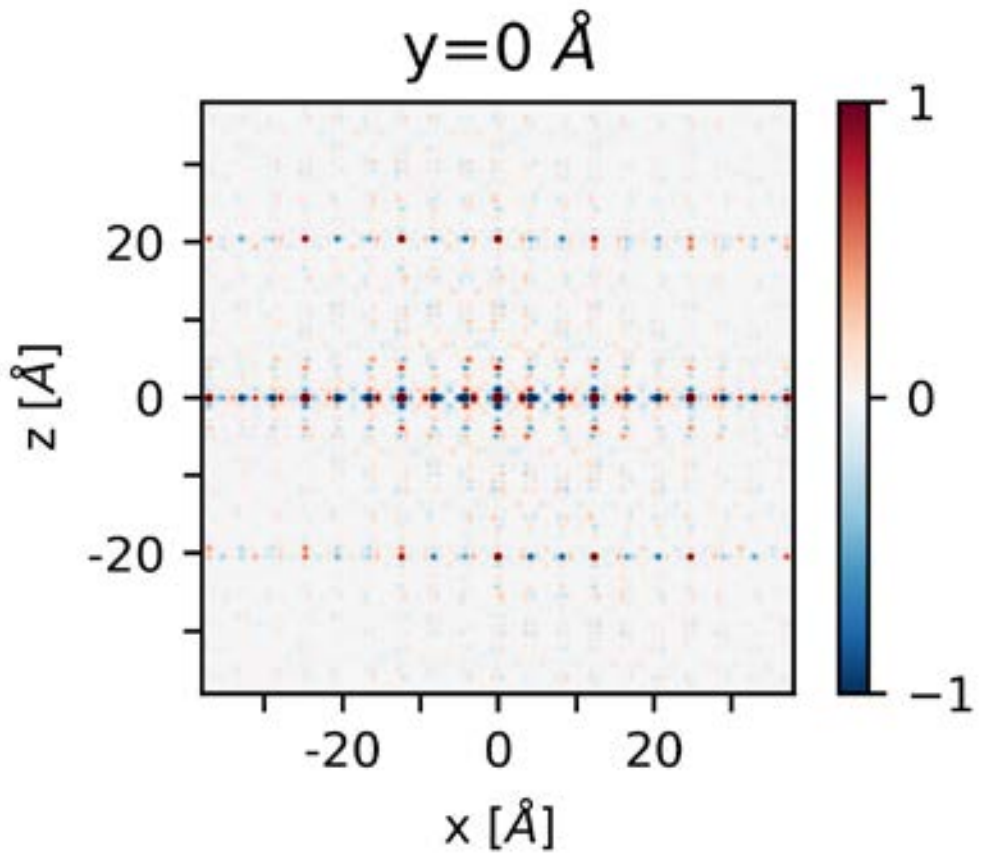

**Figure S9** Simulated 3D-ΔPDF for the model where only the mirror image disorder of the layer is used.